\documentclass[preprint,titlepage,11pt]{aastex}
\usepackage{geometry}                
\usepackage{graphicx}
\usepackage{amssymb}
\usepackage{epstopdf}
\usepackage{natbib}
\shorttitle{CHIRON TOOLS}
\shortauthors{Brewer, Giguere \& Fischer}

\title{CHIRON TOOLS: Integrated Target Submission, Scheduling and Observing Systems for a High Resolution Fiber Fed Spectrograph}

\author{John M. Brewer, Matthew Giguere, \& Debra A. Fischer}
	\affil{Department of Astronomy, Yale University}
	\affil{260 Whitney Avenue, New Haven, CT 06511, USA}
	\email{john.brewer@yale.edu}
	\email{matthew.giguere@yale.edu}
	\email{debra.fischer@yale.edu}
	
\begin{abstract}
The CHIRON spectrometer is a new high-resolution, fiber-fed instrument on the 1.5 meter telescope at Cerro Tololo Inter-America Observatory (CTIO).   To optimize use of the instrument and limited human resources, we have designed an integrated set of web applications allowing target submission, observing script planning, nightly script execution and logging, and access to reduced data by multiple users. The unified and easy to use interface has dramatically reduced the time needed to submit and schedule observations and improved the efficiency and accuracy of nightly operations. We present our experience to help astronomers and project managers who need to plan for the scope of effort required to commission a queue-scheduled facility instrument.
\end{abstract}


\date{}

\begin{document}
\maketitle

%
%
\section{Introduction}
	CHIRON is a fiber fed spectrometer with spectral resolution up to $R = 137\,000$ designed for extra-solar planet searches  \citep{2010SPIE.7735E.149S,2012SPIE.8446E..0BS,Tokovinin:2013:submitted} at the 1.5-meter telescope at Cerro Tololo Inter-America Observatory (CTIO). After initial commissioning, we found that significant effort was required to schedule the queue-based instrument for multiple users with different time sensitive requests. We also found that errors in creating and executing the nightly observing scripts resulted in time losses of up to an hour per night and $\sim 10$\% loss of data. It became evident that in addition to the planned automated data reduction pipeline, a new software interface was needed to submit targets, to generate nightly observing scripts, and to execute and log observations in order to take full advantage of the new instrument. We had significantly under-estimated the full scope of work required to commission the spectrometer; the software effort required to make this a facility instrument nearly doubled the effort required to build the instrument. 
	
	Historically, astronomers traveled to the telescope to carry out their observing programs. However, that model has changed in the past decade to more economical models where telescope operators execute observing programs or the data are acquired robotically. The 1.5-meter telescope at CTIO is part of the Small to Moderate Aperture Research Telescopes (SMARTS) consortium \citep{2010SPIE.7737E..31S} that is partly supported by the National Optical Astronomical Observatories (NOAO). Users are either awarded time through the NOAO proposal process or they purchase telescope time from SMARTS on an hourly basis. The data are generally obtained by telescope operators. During the commissioning of CHIRON, shared risk time was opened to the community and there was a large influx of users with time allocations between 5 and 100 hours over the semester. Initially, PIs e-mailed target lists to the scheduler at Yale and nightly scripts were assembled and sent as PDF files to the telescope operator at CTIO. The operator was responsible for taking all calibrations and logging any issues with observing. While still working out kinks in the instrument control system (ICS), missed observations or calibrations were frequent as were simple typos both in creating the observing script and in entering script information into the ICS and telescope control system (TCS). 
	
	We developed a software management system with the goal of improving the efficiency and scientific productivity of the 1.5-meter CHIRON queue. Our specific objectives were to reduce the time needed to build nightly scripts (with targets from multiple users and with different time sensitive scheduling requirements) and to reduce the errors in executing the scripts. The full list of goals and requirements of the software can be found in \S\ref{sec:goals_reqs}. Rapid deployment of this management system was also critical and this resulted in some software design choices that were narrowly tailored to our particular instrument and queue setup. However, the design pattern can be easily adapted for other facility instruments and we offer our code, Tools for Observing, Operating, Logging and Scheduling (TOOLS), to anyone interested in adapting it for their use.

%
%
\section{Goals and Requirements}
\label{sec:goals_reqs}
	The project goal was to improve the efficiency and scientific return of CHIRON. The following requirements were set to automate management of the CHIRON queue. 
			
	\begin{itemize}
	
	\item \textit{Rapid Deployment} \textemdash\ The project had to be completed as quickly as possible to reduce the data loss and to handle scheduling allocations for the upcoming semester.

	\item \textit{Accurate \& Complete Target Information} \textemdash\ Proper scheduling of a CHIRON observation requires specification of the target name, coordinates, brightness, slit choice, binning, calibrations (including additional observations of calibration targets), iodine cell position, number of exposures, and desired SNR and/or exposure time. E-mailed information often contained typos or incomplete information which resulted in repeated correspondence or incorrect or missed observations. 
	
	\item \textit{Simplified process for creating Nightly Observing Scripts} \textemdash\ As the number of users increased, it began taking up to two hours to schedule and format the observing scripts for each night.
	
	\item \textit{Reduction in Observing Errors} \textemdash\ The data taking GUI contained several fields that the telescope operator needed to fill for each observation. This was a time-consuming process with many failure modes. Observations were unusable if the slit, binning, I2 cell, or exposure time were not correctly set, coordinates were not entered correctly or if a standard comparison star or the nightly calibrations were skipped. If a typo was made in the object or program name, up to an hour was spent the next day trying to match observations to individual investigator programs. If coordinates were incorrectly typed into the TCS, dark time was lost tracking down this error or the wrong star was observed. 
	
	\item \textit{Detailed Observing Log} \textemdash\ Weather and early system problems also contributed to loss of observations. Communication regarding successful observations and the reasons for missed observations was a high priority and we needed a system that would not unduly burden or distract the telescope operator.
	
	\item \textit{Secure Target Information} \textemdash\ With the possibility of competing projects, it was deemed important that target information be treated as proprietary for individual programs.  Observers are only able to access information about their own programs with CHIRON TOOLS.
	
	\item \textit{Automated Data Reduction and Backup} \textemdash\ The automated reduction pipeline was integrated with our scheduling queue to save time and eliminate human handling errors. As observations are archived and spectra extracted, the queue system is automatically updated to let scientists know that they can retrieve their raw and reduced data. 
	
	\item \textit{Use Existing Hardware} \textemdash\ For both cost savings and rapid turnaround, we built the software system integrated with the existing MacOS based servers and RAID system at Yale and the Mac and Linux machines used for observing at CTIO.  To ensure that the applications would be easily portable, we kept largely to standard open source tools available on all Unix based systems.
	
	\item \textit{Intuitive Graphical User Interface} \textemdash\ An intuitive GUI was a high priority to reduce errors and training time for the tools.
		
	\item \textit{Robust to Internet Outages} \textemdash\ We needed a flexible automated system which would also tolerate interruptions in internet connectivity.
	
	\end{itemize}

%
%
\section{Application Design}

	\subsection{Hardware and Software}
		\label{sec:hardware_software}
\begin{figure}
		   \plotone{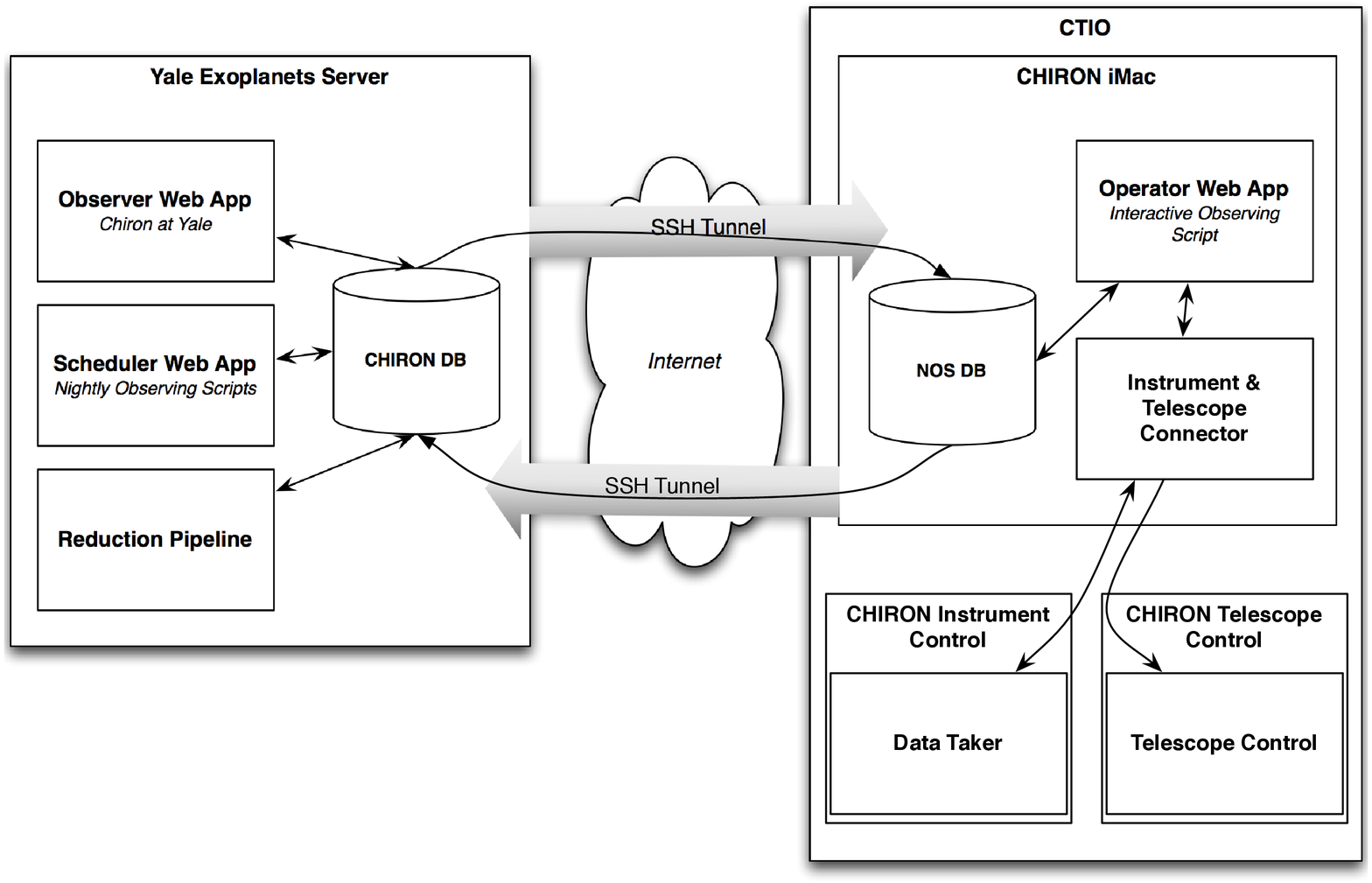} 
		   \caption{Software Overview.  The applications are divided between servers at Yale and at CTIO, with asynchronous bi-directional database syncing connecting the two locations.  The thin black arrows indicate data flow and the large shaded arrows represent the SSH tunnels which facilitate secure replication. Because the applications only communicate with the local database in each location, Internet outages do not affect nightly observing or new script creation.}
		   \label{fig:app_layout}
		\end{figure}

		The Yale Exoplanets web server is a Mac Pro running Mac OS X Server.  At CTIO, there is an iMac running the standard version of Mac OS X.  The Instrument Control System and Telescope Control Systems are Linux and VxWorks machines.  Communication between Yale and CTIO all happens between the two Mac systems.  A diagram of the system can be seen in Figure \ref{fig:app_layout}
		
		The software used to construct the web applications is a combination of PHP, SQL, HTML, CSS, Javascript, XML, JSON, and C-Shell and run on a combination of Apache2 and MySQL servers.  This is a common setup for web applications and is known as LAMP Linux or MAMP on MacOS where AMP refers to Apache, MySQL, PHP.  Bi-directional replication of a portion of the database between Yale and CTIO is handled by means of SSH Tunnels which use the Mac OS X launchd system to dynamically connect when needed and when a connection is available.  The xinetd system, prevalent on Linux, could also have been used but launched was a flexible, easy and pre-installed option on our MacOS systems.
				
	\subsection{Observer Web App}
		The first stage of development focused on a web application which allowed users with allocations for a given semester to log in and upload their target information along with specifications for the observations.  With many new principal investigators (PIs) per semester, it is important to automate account setup as much as possible.  After SMARTS time or the NOAO Time Allocation Committee (TAC) has awarded time to each observer, we use a PHP script to insert new users and add proposal and allocation information into the database.  Since all PIs are required to provide an e-mail address with their proposal we allow them to activate their account by providing their name and proposal ID and a link is e-mailed to them to allow them to set a user name and password. The PIs receive a brief tutorial that explains how to fill out the target (package) requests when they are allocated CHIRON time.
		
		Once logged in, the PI can create one or more 'plans' per semester to organize their target lists.  Adding targets is accomplished with a context sensitive form which ensures that only options pertaining to the particular type of observation and slit type are available.  Targets can be grouped together as 'packages' with calibrations or standard star observations and the total time (including approximate targeting and readout) is deducted from their allocation dynamically. Target submission has a cutoff date each semester after which the PI can no longer add or edit targets.
		
		\begin{figure}
		  \plotone{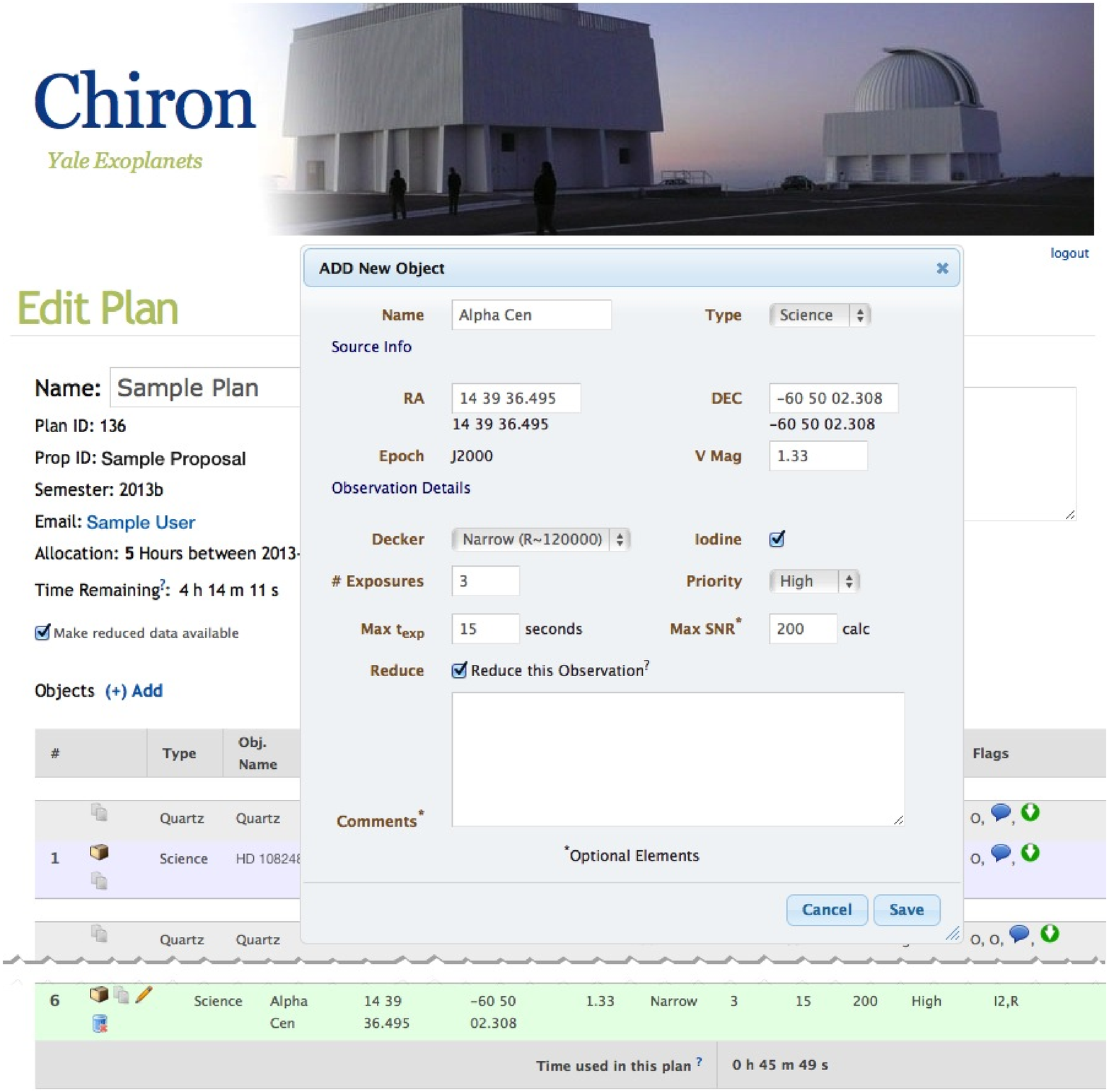} 
		   \caption{The observer web app allows investigators to quickly enter science targets and calibrations, only displaying the appropriate fields for each.  Calibration targets as well as flats can be packaged together with the associated science target simply by choosing the science target from a context sensitive drop-down menu.  As each item is saved the lines are dynamically added to the plan (line 6) and the form remains available for rapid entry.  The icons in the left column allow duplication of entire packages or just a single line item as well as edit or delete the line item.  Once a package has been observed, editing capabilities are removed (line 1) and download links give access to the reduced data.}
		   \label{fig:add_target_screen}
		\end{figure}
		
		Once scheduling is complete, the PI can monitor the observing plan to see the status of their observations and collect their data.  When targets (or packages) are added to a nightly observing script, an annotation is displayed on the PI observing plan page to show the date when the observation will occur. The annotation is automatically updated when any action occurs. If the target has been observed, a link to the raw and reduced data appears. If the target is skipped because of bad weather or mechanical failure, an auto-generated explanation appears. Importantly, all annotations are automatically saved to the PI observing plan as soon as the telescope operator selects a note from a pop-up menu on the observing script.  In addition to the annotations, users are notified via e-mail automatically once the spectra are extracted and wavelength calibrated and are ready to be downloaded.
	
	\subsection{Nightly Observing Scripts}
		The previous spectrograph on the CTIO 1.5m only had a handful of users per semester.  Nightly observing scripts were assembled in an Excel spreadsheet and exported as a PDF for the Telescope Operator.  The large increase in the number of users per semester, and per night, required a better system.
		
		When a scheduler creates a new Nightly Observing Script (NOS) for a given night, by default they are shown all targets which have not yet been observed, sorted in RA order.  Notes attached by the investigator to both their plan and individual targets are available to the scheduler for reference.  The scheduler can click on targets to add them to the script and they are immediately displayed at the top of the page along with any calibrations attached to the target.  The script displays the approximate start  time and duration of the observations based on the requested integration time, number of exposures, and acquisition and readout times.  An additional column in the script shows the scheduler if the target is visible at that point in the night and whether it is rising or setting.  They can then choose to re-organize the script by changing the sequence number of any line items.
		
		Once a script is complete, the script is sent to the Interactive Observing Script (IOS) application on the computers in Chile.  This process involves copying the necessary information for the telescope operator to records in the database which are replicated to the database at CTIO when there is an Internet connection.  If the script needs to be updated at any point, it can be re-sent with each submission resulting in a separately versioned script.
		
		Writing good scheduling software is complex, time consuming, and still usually requires human intervention.  The NOS tool has allowed us to eliminate most of the scheduling time without spending a lot on software development.  Since both target and script information are stored in the database, it will be possible to add automated scheduling software at a later date and still retain the NOS interface for human tweaking of automatically generated schedules.
		
	\subsection{Interactive Observing Script}
	\label{sec:IOS}
  We developed quality control pages to monitor the nightly observations and found that on average, 40\% of the observations had an error in the FITS header information. In minor cases, such as a typo in the object name, we could correct the FITS headers. However, correcting FITS headers for the observing queue was outside the scope of work that we could commit for CHIRON. For more significant errors, such as an incorrect slit setting, binning, coordinates, I2 cell or exposure time, the data might not even be useable by the PI. Therefore, we developed an interactive observing script (IOS) for use by the telescope operator. This web-based app simplified observing for the telescope operator by sending target coordinates to the TCS and auto-filling all fields in the data-taking GUI. 
		
		We defined an interface and a set of requirements for the IOS web application as well as an application programming interface (API) for a piece of middleware, the Instrument and Telescope Connector which could take command line requests and return XML responses. This allowed the web application developer (Methanie Binder) to develop to the API without knowledge of how the telescope or instrument control software worked.  It also allowed later additions such as integration with the exposure meter to take place solely within the middleware without having to call back the web developer.
				
		When a nightly observing script is completed (typically at Yale by the CHIRON queue manager), targets are locked from further changes in the PI observing plan and the NOS is copied into a smaller table structure to reduce the data flow for replication at CTIO. The interactive observing script is auto-generated from the NOS for the telescope operator and includes the formatted list of targets and calibrations with active links to any extended notes provided by the PI when scheduling a target package. Occasional Target of Opportunity (ToO) requests for a different instrument can also be added to the script.
		
		Each line in the IOS has buttons allowing the operator to send or skip the observation or calibration.  Selecting a line item sends all relevant information to the middleware which in turn updates the data taker and (if it is a target) the telescope coordinates.  Once a line item has been sent, it is possible to 'resend' in the event of an error. A pop-up menu is included for each line of the IOS so that the telescope operator can quickly tag the reason (weather, instrument failure, other) for skipping particular observations. If the internet connection is active, any information recorded by the telescope operator is automatically replicated back to the master CHIRON database and appears immediately on the PI's observing plan. If the internet connection is down, the information will be replicated as soon as the connection returns.
		
	\subsection{Integration with Instrument and Telescope Controls}
		As discussed in \S\ref{sec:IOS}, the IOS web application does not attempt to talk directly with the instrument control system (ICS) or the telescope control system (TCS), but instead communicates via a command-line tool which returns XML responses.  This segregation avoids the need for a web programmer to make updates when changes are made to the controller software or to the machines that host the ICS and TCS. 
		
		Both the ICS and TCS live on separate machines but do not currently have network sockets listening for commands.  Instead, tasks such as sending coordinates to the telescope or updating settings in the data taker must run through command line tools on the respective machines.  The middleware shell script handles the network communication and encapsulates the syntax for both the ICS and TCS tools.  The API consists of a single command, send\_object, with arguments consisting of all information required for both systems and whether the object is a science target or a calibration.
		
		The middleware is launched by the web app and executes remote commands on the ICS and TCS computers as appropriate. Status information is returned to standard out which is read by the web app and returned to the browser.  The script continues to monitor the progress of the setting changes every second and writes log information to a file, which can be reviewed later. A status window is launched so that the telescope operator can see when all of the changes are complete.  Though the middleware is capable of monitoring the exposures, it currently only returns status information to the web application about the success of setting the ICS and TCS items. 
		
	\subsection{Database Design and Replication}
		\label{sec:db_design}
		\begin{figure}
		  \plotone{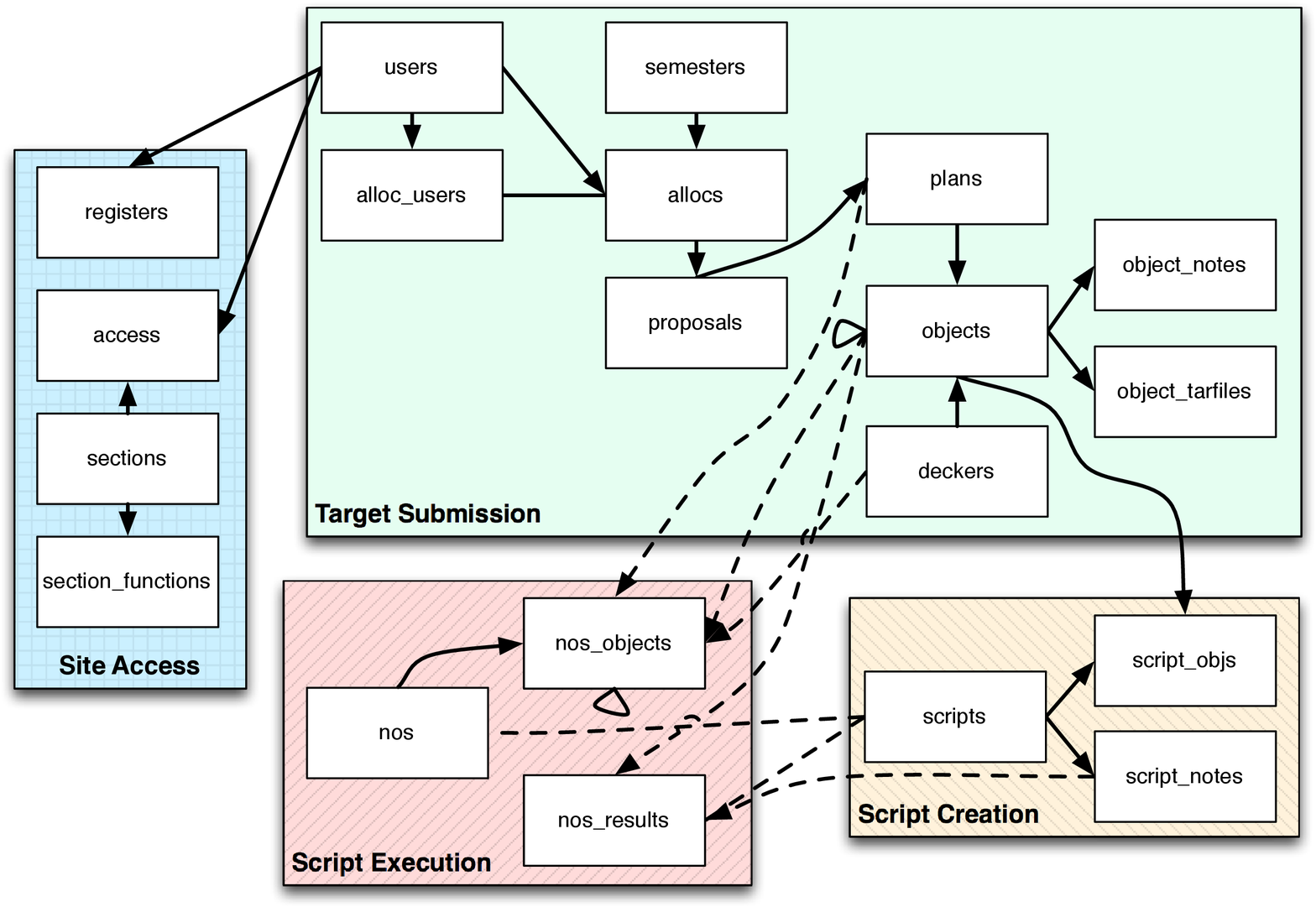} 
		   \caption{A schematic view of the Chiron database ('Chiron DB' in Figure \ref{fig:app_layout}).  Solid lines show relational links between tables and dashed lines represent links where no constraints are enforced.  Lines with arrows show the direction of one-to-many relationships and loops indicate self-references.  The boxes group logical portions of the application.  Most of the database structure is generic to any similar effort and could be directly re-used.  Tables which might need to be modified are those which store instrument specific settings: \emph{objects}, \emph{nos\_objects}, and \emph{deckers} (spectrograph slit and fiber options), though another high resolution spectrograph should be able to use the database structure unaltered.  The 'Script Execution' unit is almost a self-contained database meant to reduce replication requirements and any changes to the \emph{objects} table would need corresponding changes to the \emph{nos\_objects} table.}
		   \label{fig:db_layout}
		\end{figure}
		
		We designed a relational database to manage all of the information for targets, scripts, logging, and access to the applications. The database can be viewed as separate logical units (Figure \ref{fig:db_layout}). With the exception of the \textit{Script Execution} unit, it was designed as a single normalized\citep{Codd:1970:RMD:362384.362685} set of tables.  The structure and fields are generic to any queue based astronomical observing with time allocation blocks assigned by semester.  Information specific to the instrument is contained within the \textit{object} and \textit{decker} (slit and fiber options) tables and the related \textit{nos\_objects} table.
		
		Due to sporadic outages and slow-downs of the Internet connection to the 1.5 meter telescope at CTIO and to avoid impacting the mountain network during observing times, we attempted to limit the traffic needed to link the database on the mountain to the one at Yale.  The nightly observing script tables (\textit{nos} and \textit{nos\_objects}) duplicate all information for scheduled targets contained in the main scheduling tables (\textit{scripts} and \textit{script\_objs}) as well as necessary information from linked tables (such as \textit{objects} and \textit{deckers}) in order to reduce the amount of information replicated to CTIO.  The simplified structure has the side-benefit of obscuring proprietary observer information at the telescope and includes date-stamps and version information for reference.
		
		Replication of the \textit{Site Access} and \textit{Script Execution} table groups along with the \textit{users} table is accomplished using the built-in MySQL replication over secure SSH tunnels as described in \S\ref{sec:hardware_software}.  Most of the tables are read-only at CTIO which eased the replication setup.  Only the results from the nightly script (\textit{nos\_results} table) can be updated in both locations; this is the only table participating in bi-directional replication.  The others are all synced from Yale to CTIO and generally have very low update frequencies.
		
		During an observing run, the IOS application used by the telescope operator reads and writes to the local database on the iMac at CTIO.  Log entries are saved to the \textit{nos\_results} table which is replicated on-demand when an Internet connection is present and typically consists of less than a few hundred bytes of data per observation.
	
	\subsection{Reduction Pipeline}
	To take advantage of the lower demand on intercontinental bandwidth at night, data are rsynced from CTIO to Yale on an hourly basis. This allows processing, analysis and distribution to take place the next morning. Data are then processed in IDL using a modified version of the \textit{REDUCE} package created by Piskunov \& Valenti \citep{2002A&A...385.1095P}. Both the raw and extracted and wavelength calibrated reduced files are in the FITS standard \citep{2010A&A...524A..42P}. Files are packaged and compressed into tarfiles and their locations are entered into the CHIRON MySQL database at the end of the pipeline. When users log in to their CHIRON accounts, hyperlinks to the combined raw and reduced tar files appear in each target row, allowing users to download their data.
		
%
%
\section{Conclusions}
	
	\subsection{Reduced Time Requirements}
		One of the primary reasons for spending the time to develop this software was to reduce the amount of time we were spending on the mundane tasks of observation planning and distributing the data.  Initially, construction of nightly observing scripts required up to two hours per night including time spent communicating with observers.  Schedulers now see a list of unobserved targets and can click to add them to and re-order them in the script which shows target elevation and time available.  The PI portion of the application has already ensured that all necessary data is in the database.  After full implementation of the software, script preparation has been reduced to $\sim$15 minutes per night.  The system has been easy to maintain, requiring little intervention other than operating system updates.
		
		We have no data on how much time PIs spent exporting their observing lists before the software upgrade. However, we know that $\sim$10\% of the data was unusable and repeat observations were required. There have been a few requests for features which would make target submission smoother such as the ability to duplicate observations or entire plans and the ability to upload basic target information (name, RA, DEC, V mag) in bulk.  These are sensible upgrades but there is no funding for this effort. 
		
	\subsection{Improved Science Output}
		The second reason for developing this software was to improve the efficiency and science return of the instrument.  Before implementing this software, up to 40\% of the observations had some type of error in the FITS header; roughly 10\% of data was unusable because of incorrect settings manually entered into the data-taking GUI or because calibration targets were missed; manually typing instrument control settings in the GUI and target coordinates for the TCS took several minutes for every target or up to an hour per night. The new software system has eliminated the instrument setup errors (any errors that exist are due to incorrect target requests by the PI) and there are no longer any lost observations due to missed calibrations. Thus, we estimate that we have recovered nearly two hours of observing time every night. 
	
	\subsection{Lessons learned}
		It is worth restating the obvious: in developing a facility instrument, the task of organizing observations, data reduction pipelines and distributing reduced data is a significant effort.  We estimate that the time to develop the scheduling software (including the PI user interface, the NOS, the middleware, the IOS) totaled to 6 month FTE effort. This was all an effort that we did not plan for when proposing to build CHIRON. The algorithms provided in this paper can help in planning the scope of software integration.
	
		Since the previous spectrograph was thought to have adequate software, the need for significantly revised software was not recognized until commissioning the instrument.  In order to fully exploit the capabilities of a new instrument it is critical to evaluate the entire process from proposal through delivery of reduced data.  Identifying and planning for efficiency bottlenecks in the process can be critical to the success of the instrument and should be included in the scope of work for the hardware.

%
%
\section{Summary}
	While commissioning the CHIRON spectrograph at CTIO, challenges with the target collection, nightly planning, and data distribution were identified which led to the development of this suite of software applications.  Overall instrument performance was greatly improved by a development effort which was relatively modest in comparison to time spent on the instrument design and construction itself.  However, analysis of the new use cases of the instrument prior to commissioning would have further reduced development time and made for smoother commissioning.
	
	We share our experience in building a facility-class software interface and offer our software as a template on which to build similar target submission, scheduling, and observing systems for facility instruments.  Due to the focus on rapid deployment, parts of our software are necessarily tied closely to CHIRON and the specifics of the Yale and CTIO systems.  The software is available upon request, as-is, but could be adapted to other instruments and settings.  To facilitate this, we have carefully segregated instrument specific items in the database and used middleware to abstract the communication of object information to the instrument.

	\acknowledgements
	Thanks are due to Marco Bonati for making the data taker remotely accessible and Andrei Tokovinin for helpful discussions and feedback.  We thank Michele Beleu for keeping the project on schedule, and Methanie Binder for rapid turnaround on the web design.  We also gratefully acknowledge the support of NASA grant NNX12AC01G and NSF grant AST1109727.

\bibliography{ms}

\end{document}